\documentstyle[prl,aps,twocolumn]{revtex}

\newcommand{\OO}{{\cal O}}
\newcommand{\cf}{{\em cf. }}
\newcommand{\ie}{{\em i.e.}}

\newcommand{\rhs}{{\em rhs }}
\newcommand{\re}{{\rm Re\,}}
\newcommand{\im}{{\rm Im\,}}
\newcommand{\aleq}{\stackrel{<}{\scriptstyle\sim}}

\begin{document}



\author{F. Bentosela$^{1}$, P. Exner$^{2}$, V.A. Zagrebnov$^{1}$}

\title{A Mechanism of Porous--Silicon Luminiscence}
\address{$^1$ Universit{\'e} de la M\'editerran\'ee (Aix--Marseille II), 
and Centre de 
Physique Th\'eorique, CNRS, Luminy, Case 907,  
F--13288 Marseille; 
{\it bentosela@cpt.univ--mrs.fr, zagrebnov@cpt.univ--mrs.fr} \\
$^2$ Nuclear Physics Institute, Academy of Sciences,
CZ--25068 \v Re\v z, and  Doppler Institute, Czech Technical University,
CZ--11519 Prague; 
{\it exner@ujf.cas.cz}} 

\date{\today}

\maketitle

\begin{abstract}
We discuss the discrete spectrum induced by bulges on threadlike 
mesoscopic 
objects, using two models, a continuous hard--wall waveguide and a 
discrete 
tight--binding model with two sorts of atomic orbitals. We show that 
elongated 
bulges induce numerous quasibound states. In the discrete model we also 
evaluate the probability of transition between the localized states and 
extended ones of the ``valence" band. We suggest this as a mechanism 
governing the porous--silicon luminiscence. In addition, the model 
reproduces 
the dominance of nonradiative transitions, blue shift for finer textures 
and 
luminiscence suppression at low temperatures.
\end{abstract}


\vspace{5mm}

\vspace{5mm}

\noindent
The effect of luminiscence of porous silicons has attracted a lot of
attention recently \cite{silicon}. There are various attempts to explain 

it, but none of them can be regarded as fully convincing at present. It 
is 
clear that the porous material texture plays the decisive role, because 
first the effect is absent in the bulk, and second, a refinement of the
structure is known to cause a blue shift of the emitted light.
In this Letter we intend to discuss one possible quantum mechanical 
mechanism which employs transitions between the valence band and a 
large family of localized states below the conductance bend; we put 
emphasis on describing the geometric conditions under which such families 

may exist. 

It has been suggested that quasibound states in small crystallites
may play important role \cite{herino}. It is natural to expect that the 
interior of the porous medium resembles a sort of a calcite cave 
containing 
not only loose--end material ``drops'' but also other structures; our 
main 
hypothesis is that {\em a significant portion of them are threadlike 
objects 
of a varying cross section.} Under this assumption we may employ recent 
results on electron bound states in quantum wires which are bent, 
protruded, 
or coupled laterally to another wire 
\cite{bent qwg,protruded qwg,protruded2,coupled qwg}. The mechanism 
behind 
the existence of these bound states is an effective attractive potential 

induced by the geometric modification of the tube. Our key observation is 

that if the deformation extends over a long interval (relative to the 
tube 
cross section), the waveguide can support numerous bound states and the 
discrete spectrum has typical one--dimensional features: most eigenvalues 
are 
found at the bottom of the spectrum, \ie, away of the continuum. Hence if 

the variation of the tube cross section produces protrusions which are 
rather 
long than wide, such a tube has many more quasibound states than other 
conceivable structures, so the corresponding radiative transitions are 
responsible for the most part of the emitted light.

Below we shall illustrate this feature on a tube with a single elongated 

bulge. On the other hand, any model of porous--silicon luminiscence has 
to be able to reproduce the other experimentally established properties,
notably the dominance of the nonradiative transition mode as well as the
frequency and temperature dependence of the effect. For this purpose 
the free--particle quantum waveguide model is oversimplified, because 
its continuous spectrum consists of a single band. This motivates us to 
treat the essentially same situation in the tight--binding setting, 
considering chains of ``atoms" to which other chains of finite  length 
are 
laterally attached. If the atomic orbitals are of two different sorts, 
the 
spectrum  of an infinite chain can consist of distinguished bands which 
would play the role  of the valence and conductance band, respectively.  

Adding a finite chain  will cause appearance of bound states whose 
distance 
from the band edges  is controlled by the coupling strength between the 
two 
chains.

Truncating the discrete ``tube", we are able to find the spectrum and the 

corresponding eigenfunctions numerically. This will allow us to estimate 

the rate of transition between the quasibound states below the 
conductance 
band and extended states in the valence band. This quantity can be 
compared to the probability of nonradiative transitions due to a 
tunneling 
escape of an electron localized in a bulge to a neighbouring 
bulge or to the bulk from which the treadlike structure spreads. 

Let us describe briefly the two models; more details will be given in a 
forthcoming paper \cite{bez}. In the continuous model we consider a tube 

with hard walls which has a constant cross section except for 
a finite part where it is protruded \cite{units}. The bulge produces 
bound states no matter how small it is \cite{protruded2}, but of course,
the number of such states and the distribution of the corresponding
energy levels depend substantially on the geometry. For instance, a
hard--wall planar strip of a unit width with a stub of the same width
and length $\,\ell\,$ considered in Ref.\cite{stub} has just one bound
state $\,\lambda(\ell)\,$ such that $\,\lambda(\ell)= \pi^2\!
-\pi^4\ell^2\! +\OO(\ell^4)\,$ for small $\,\ell\,$ and
$\,\lim_{\ell\to\infty} \lambda(\ell)\le 0.93 \pi^2\,$ ---
\cf \cite{protruded qwg,ess}; making the protrusion two--sided, we have 
still one bound state with the eigenvalue which cannot be lower 
than $\,0.66\pi^2\,$.

On the other hand, elongated bulges produce numerous bound states. As a 
simple example,  consider a boxlike protrusion on a straight planar 
strip, 
so the width is $\,1\!+\!\eta\,$ on an interval of a length $\,L\,$ and 
one
otherwise. By a bracketing argument \cite{bracketing} the discrete
energy levels are squeezed between the eigenvalues of the Laplacian on 
the rectangle $\,[0,L]\times [0,1\!+\!\eta]\,$ with the Dirichlet 
condition 
on the ``parallel" boundary and Dirichlet or Neumann, respectively, on 
the
``perpendicular" one, that is,
   \begin{equation} \label{boxlike bulge}
\left({\pi j\over{1+\eta}} \right)^2+ \left({\pi(n\!-\!1)\over L}
\right)^2 \le\, \lambda_{j,n}\le\,
\left({\pi j\over{1+\eta}} \right)^2+ \left({\pi n\over L}
\right)^2
   \end{equation}
for $\,n=1,2,\dots\,$. The discrete spectrum consists of those
$\,\lambda_{j,n}\,$ which are below $\,\pi^2$, the bottom of the
continuous spectrum; it is clear that with the lowest transverse
mode, $\,j=1\,$, such states exist for any $\,\eta>0\,$ as long as
$\,L\,$ is large enough. Moreover, in the case $\,L\gg 1\,$ there are
numerous bound states, with most eigenvalues being concentrated in
the vicinity of $\,\pi^2(1\!+\!\eta)^{-2}$, or the higher thresholds
$\,(\pi j)^2(1\!+\!\eta)^{-2}$, provided the latter are below the
bottom of the continuous spectrum. These conclusions extend easily to
a tube with a steplike bulge in three dimensions.

The fact that elongated bulges produce many bound states is not
restricted to the above simple example; on the other hand, the
eigenvalue distribution depends substantially on the protrusion
shape. To get a better understanding, consider a tube whose cross
section $\,\Sigma_x\,$ is constant for $\,|x|>\,{1\over 2}L\,$ and
varies smoothly in the interval $\,\left\lbrack-{1\over 2}L, {1\over
2}L\right\rbrack\;$ (see Fig.1a). 
For a fixed $\,x\,$ let $\,\nu_1(x)< \nu_2(x) \le
\nu_3(x)\le\cdots\,$ denote the eigenvalues of the Laplacian with the
Dirichlet condition in $\,L^2(\Sigma_x)\,$; the corresponding
eigenfunctions are $\,\chi_j(x,y)\,,\; j=1,2,\dots\,$; $\;y\,$
being the transverse variable(s). The ``full" wave function may be
then written in the form $\,\psi(x,y)= \sum_j a_j(x)\psi_j(x,y)\,$
with the normalization $\,\int_{-L/2}^{L/2} \sum_j |a_j(x)|^2 dx =1\,$.

The protrusion--induced discrete spectrum is essentially determined
again by the spectrum of the bubble alone; one can employ the
bracketing argument closing the bulge at $\,x=\pm {1\over 2}L\,$ by
the Dirichlet and Neumann ``lid", respectively. 
If  we assume now that the bulge is {\em long} and its cross section 
{\em changes only slowly} with respect to $\,x\,$ the longitudinal
derivatives of $\,\chi_j\,$ may be neglected and we arrive at an
Born--Oppenheimer type approximation: the stationary Schr\"odinger
equation decouples into a family of equations for the slow motion,
   \begin{equation} \label{BO approximation}
-a''_j(x) +\nu_j(x)a_j(x)\,=\, Ea_j(x)\,,
   \end{equation}
where the the transverse eigenvalues play role of the potentials. 
At the same time, if the bulge is long the eigenvalues $\,E_{j,n}\,$
of the $\,j$--th equation are determined approximately by the
semiclassical quantization condition
   \begin{equation} \label{sc quantization}
\int_{M_j(E)} \sqrt{E\!-\!\nu_j(x)}\,dx \,=\, n\pi+\mu_j\,,
   \end{equation}
where $\,M_j(E):= \{\,x:\: \nu_j(x)\le E\,\}\,$ is the classically
allowed region; the explicit value of the Maslov factor $\,\mu_j\,$ is 
not important as long as we are interested in the distance $\,\delta
E_{j,n}=E_{j,n+1}\!-E_{j,n}\,$  between the adjacent energy levels
which determines the density of states $\,\rho(E)\,$. Expanding the
square root and neglecting the difference between $\,M_j(E_{j,n})\,$
and $\,M_j(E_{j,n+1})\,$, we find that the latter approaches in the
limit $\,L\to\infty\,$ the form
   \begin{equation} \label{density of states}
\rho(E)\,=\, {1\over 2\pi}\, \sum_j\, \int_{M_j(E)}\, {dx\over
\sqrt{E\!-\!\nu_j(x)}} \;;
   \end{equation}
recall that we are interested only in the behavior of this
function below $\,\nu_1(L/2)\,$, the bottom of the continuous
spectrum, where just one or several lowest transverse modes can have
$\,M_j(E)\ne \emptyset\,$.

Returning to our example of a boxlike bulge on a unit--width strip, we
find that for large $\,L\,$ the $\,j$--th mode contribution to the
discrete--spectrum density is 
   \begin{equation} \label{rho box}
\rho_j(E)= \left(L\over 2\pi\right)\, \left\lbrack
E-\left(\pi j\over 1+\eta\right)^2 \right\rbrack^{-1/2}
   \end{equation}
with a singularity at $\,E_j^{min}:= \left(\pi j\over 1+\eta\right)^2\,$. 

Other shapes may change the form of the distribution; it is more 
concentrated close to the bottom of the discrete spectrum the closer is 
the bulge to the cylindrical shape. For instance, consider the strip of 
the width $\,d\left(x\over L\right)\,$, where $\,d(\xi):=
(1\!+\!\eta) (1\!-\!b\xi^2)^{1/2}\,$ and $\,b\,$ is chosen in such a
way that $\,d(\pm 1/2)=1\,$. The $\,j$--th term on the \rhs of
(\ref{density of states}) is then expressed as
   \begin{equation} \label{example 1}
\rho_j(E)\,=\, {L(1\!+\!\eta)\over 2\pi\sqrt{\eta(2\!+\!\eta)}}\; {\rm
E}\left(\sqrt{E\!-\!E_j^{min}\over E}\, \right)\,,
   \end{equation}
where $\,E_j^{min}\,$ is the same as above and$\,{\rm E}\,$ is the full 
elliptic integral of the second kind \cite{as}; it has still a peak at 
$\,E=E_j^{min}\,$ but less pronounced.

Let us now pass to description of the tight--binding model. We employ
the simplest possible choice for the atomic geometry as well as for 
the interactions between orbitals. We consider $\,N\,$ parallel chains of

atoms forming a strip in the plane to which we add $\,M\,$ finite--length 

chains which constitute a bulge (Fig.1b). To mimick the band structure of 

the semiconductor spectrum, one can choose interactions between orbitals 

(side--diagonal elements of the tight--binding Hamiltonian) switching 
between two values $\,{\bf a}\,$ and $\,{\bf b}\,$ in the horizontal  
direction; vertically one can choose the same structure or simply a 
single
coupling constant $\,c\,$.  

If one has an infinite horizontal strip of width $\,N\,$ with no bulge
the corresponding spectrum can be obtained summing the spectrum of one 
horizontal infinite chain (\ie, the pair of intervals 
$\,(-{\bf a}\!-\!{\bf b},{\bf b}\!-\!{\bf a})\,$ and 
$\,({\bf a}\!-\!{\bf b},{\bf a}\!+\!{\bf b})$) and the discrete 
spectrum corresponding to a vertical line of $\,N\,$ atoms; the latter
is of course contained in the mentioned intervals if the structure is
the same in both directions. The resulting spectrum still exhibit gaps
if $\,{\bf a}\,$ and $\,{\bf b}\,$ are chosen appropriately; in general
they become narrower with increasing $\,N\,$.

The spectrum of an infinite strip of width $\,N\,$ with a finite number 
of 
bulges of width $\,M\,$ has a continuous part identical with that of the 

``unperturbed" strip and eigenvalues outside of it. The latter are 
nevertheless 
contained in the  spectrum of a strip of width $\,N\!+\!M\,$. Fig.2 shows 
the
eigenvalue 
plot obtained numerically for a chain ($\,N\!=\!1\,$) of 40 ``atoms" and 
a bulge 
of 14 ``atoms", $\,a\!=\!3,\, b\!=\!1\;$ (in the vertical direction 
$\,b\!=\!1\,$). We can  distinguish the eigenvalues in the intervals 
$\,(-4,-2)\,$ 
and $\,(2,4)\,$  corresponding to the extended states of the ``valence" 
and 
``conduction" bands, and those outside corresponding to states localized 
mainly on the
bulges with an exponential  decay outside. In case of several bulges it 
may occur that an
eigenstate is  supported by more than one of them; this happens typically 
if the system 
has a symmetry. Notice that the extended states are not Bloch states due 
to the lack of
translational invariance.  

The knowledge of the eigenfunctions makes it possible to compute the 
radiative transition probability between the excited bound states living 
in 
the bulges and the valence--band extended states which is given in 
general 
by the Fermi golden rule,
   \begin{eqnarray} 
W_r(\omega)\! &=& \! {2\over 3}\:
{e^2\over  4\pi\epsilon_0 \hbar c}\: {\omega^3\over c^2}\: {1\over V}\: 
\sum_{i,f}\,\delta (E_i\!-\!E_f\!-\!\hbar\omega)\nonumber 
\\ && \times \left| \int_\Lambda (\vec e.\vec r)\psi_i(\vec r)
\psi_f^*(\vec r)\,d^3x\, \right|^2 \,. \label{radiative tp}
   \end{eqnarray} 
We have evaluated the matrix element in question. It is nonzero but not 
large; the value is typically at least $\,2-3\,$ orders of magnitude 
below 
the upper bound given by the potential step between the bulge ends.

Inserting the values of the constants into (\ref{radiative tp}), the 
above 
observation tells us that the transition probability does not exceed 
$\,10^8\, s^{-1}$; it increases, but not more than one order of 
magnitude,
when $\,\omega\,$ runs through the visible spectrum. The last named 
property   
conforms with the experimentally observed shorter lifetime at the blue 
edge 
of the spectrum \cite{silicon}.

It is further known \cite{silicon} that $\,W_r(\omega)\,$ exhibits a 
dramatic 
decrease below the room temperature. To explain this effect one has to 
take 
into account that the final--state probability is determined by the Fermi 

distribution, and therefore the the matrix element in (\ref{radiative 
tp}) 
should be multiplied by $\,P_{\beta,\mu}:= 1\!-\!\left( 
e^{\beta(E_F-\mu)} 
+1\right)^{-1}$. Assuming that the chemical potential takes value in the 
middle 
of the gap between the two bands, the above factor is of order of 
$\,e^{-40}\,$
at the room temperature and the decrement is inversely proportional to 
$\,T\,$;
the suppression is larger at the blue edge of the spectrum.

Other properties of this model also conform with experience for the 
effect under consideration. Long bulges support many excited states which 
is 
necessary to create a macroscopic luminosity output. At the same time,  
a simple scaling argument shows that the distance between the bound 
states 
and the valence band increases as the lateral size of the tubes and 
bulges 
become smaller; hence a finer material texture results in a blue shift.

Experimental data show a low emmision efficiency of photoluminiscence 
measured at room temperature. This strongly suggests that the radiative 
recombination $\,W_r\,$ is dominated by the nonradiative probability 
$\,W_{nr}\,$ which involves the escape of the confined carriers 
(electrons--holes) from a bulge into a more extended/less passivated 
neighborhood 
where a nonradiative recombination can occur. Hence the  emitted 
intensity
$\,I(\omega)\sim W_r(\omega)\tau(\omega)\,$, where the lifetime 
$\,\tau(\omega)=
\left(W_r(\omega)+W_{nr}(\omega) \right)^{-1}$. Independent measurements 

\cite{silicon} of $\,I(\omega)\,$ and 
$\,\tau(\omega)\,$ show that $\,W_{nr}\gg W_r\,$ and $\,W_{nr}(\omega)= 
A\, e^{\alpha\hbar\omega}\,$. In the framework of both our model a decay 
process
related to $\,W_{nr}\,$ occurs if the bulged tube is connected to a wider 
part 
of the structure (bulk). The tunneling probability can be estimated in 
the 
second model from the eigenfunction decay \cite{tunneling}:
   \begin{eqnarray}
W_{nr}(\omega) &=& \im(E_i\!-\!E_f)/\hbar \nonumber \\
&=&{1\over \sqrt{2m_e}}\, \sqrt{\re(E_i\!-\!E_f)}\: 
|\psi_{\re E_i}(L)|^2\,, \label{escape}
   \end{eqnarray}
where $\,L\,$ is the tunneling distance. For typical light photon 
energies we 
get $\,W_{nr}\sim 10^{5}\,e^{-\gamma(E) L}\,$, where $\,\gamma(E)\,$ is a 
function 
of the distance between the eigenvalue and the bottom of the 
``conduction" band. 
We get $\,W_{nr}\gg W_r\,$ at the room temperature as long as $\,L\aleq 
50\, 
{\rm a.u.}\,$; for a cooler material and bluer light the dominance is 
preserved at
longer distances.

It is certainly not easy to decide which mechanism is responsible for
the porous--silicon luminiscence as long as we know little about the
actual texture, and it is fully conceivable that the effect comes from 
conspiracy of different physical processes. On the other hand, it seems 
to be straightforward to check experimentally whether the states 
discussed in this letter may contribute, since quantum wires with bulges 

of appropriate shape can be fabricated. One could, {\em a fortiori,} 
taylor 
in this way luminiscent systems emitting light of prescribed properties.
Moreover, since the mechanism producing bound states in infinite tubes
are similar, the same can be done for quantum wires with numerous bends, 

or pairs of wires coupled laterally through a long ``window".

In conclusion, we have presented a mechanism which could be responsible 
for 
the porous--silicon luminiscence illustrating it on two models. Despite 
the 
simplifications, they yield the basic features, \ie, the existence of 
numerous quasibound states away of the continuum, the dominance of 
nonradiative 
transitions and the spectral shift associated with refining the texture.

\subsection*{Acknowledgments}

Enlightening discussions with F.~Arnaud d'Avitaya, I.~Berbeziet,  
J.~Derrien, and L.~Vervoort are gratefully acknowledged. P.E. thanks 
Centre 
de Physique Th\'eorique, C.N.R.S., where this was work was done for the 
hospitality extended to him. The research has been partially supported by 

the Grant AS CR No.148409.

\vspace{10mm}

\subsection*{Figure captions}

{\bf Figure 1} \quad The models. (a) A tubular guide with a bulge. The
bound states of an infinite tube change to quasibound when we couple it
to the bulk. (b) The tight--binding model.
\\ 
{\bf Figure 2} \quad The spectrum of the tight--binding model.


\begin{thebibliography}{article}
\bibitem{units}
For the sake of simplicity we put $\,\hbar= 2m=1\,$ when explaining 
the mechanism.
\end{thebibliography}
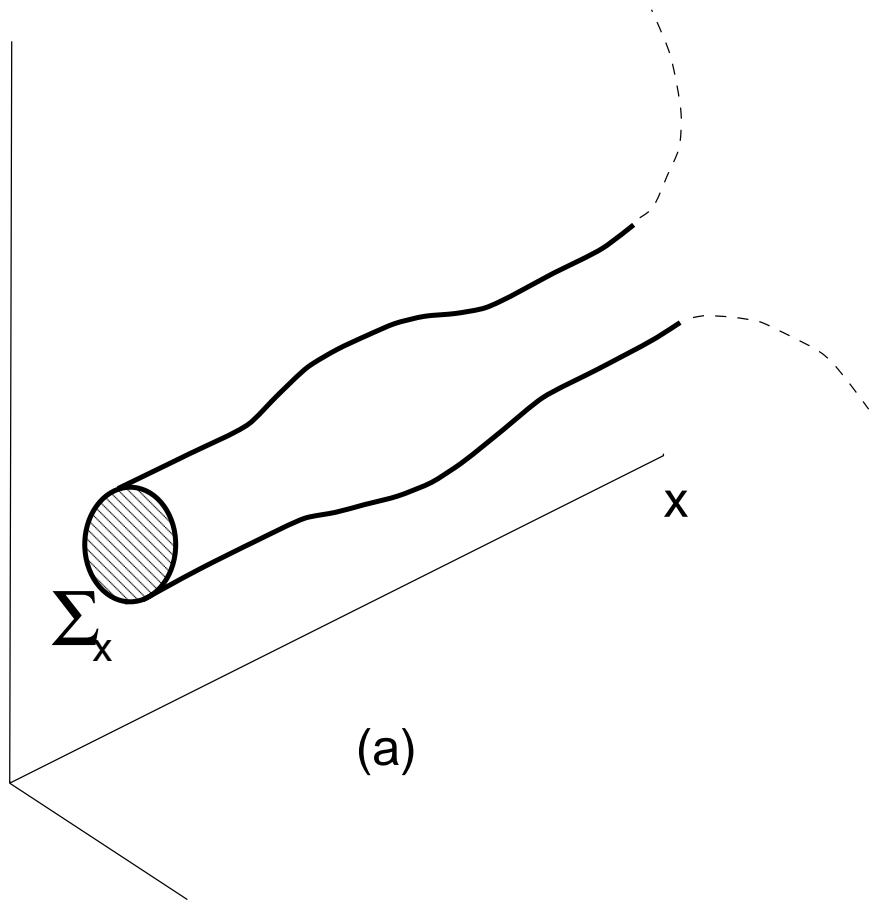
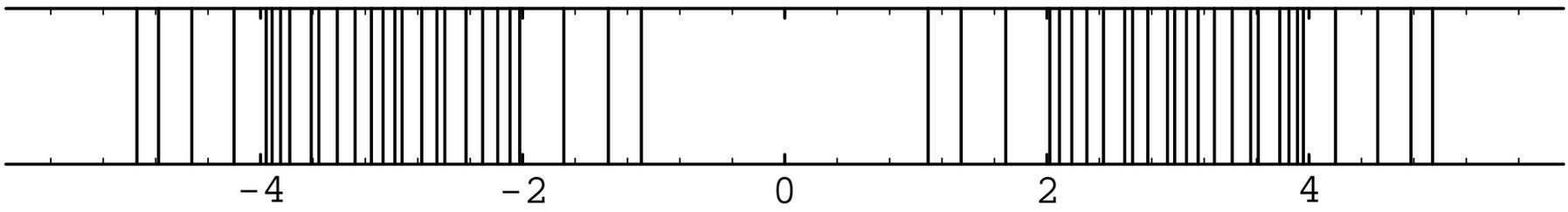
\end{document}